\documentclass[12pt,thmsa]{article}
\usepackage{amssymb}
\usepackage{graphicx}

\makeatletter
\newcommand{\row}[1]{\mathord{\buildrel{\lower3pt\hbox{$\scriptscriptstyle\rightarrow$}}\over #1}}

\newcommand{\dyadic}[1]{\mathord{\dyadic@rrow{#1}}}
\newcommand{\dyadic@rrow}[1]{
\begin{picture}(12,12)(-1,0)
\put(-2,10){\makebox(0,0)[t]{$\scriptscriptstyle\downarrow$}}
\put(-2,11){\makebox(0,0)[l]{$\scriptscriptstyle\longrightarrow$}}
\put(5,0){\makebox(0,0)[b]{$#1$}}
\end{picture}
}

\newcommand{\ket}[1]{\bigl| #1 \bigr\rangle}

\input{tcilatex}
\begin{document}

\begin{center}
\textbf{Dynamics of Bloch vectors and the channel capacity of a non
identical charged qubit pair}

N. Metwally, M. Abdel-Aty and A.-S. F. Obada$^{*}$
\\[0pt]
Mathematics Department, College of Science, Bahrain University,
32038 Bahrain
\\
$^*$Mathematics Department, Faculty of Science, Al-Azhar University,
Cairo, Egypt
\end{center}

We have considered a system of two superconducting charge qubits
capacitively coupled to a microwave resonator. The dynamics of the
Bloch vectors are investigated for different regimes. By means of
the  Bloch vectors and cross dyadic we quantify the degree of
entanglement contained in the generated entangled state. We consider
different values of the system parameters to discuss the dynamics of
the channel capacity between the qubits. We show that there is an
important role played by initial state settings, coupling constant
and the mean photon number on generating entangled state with high
degree of entanglement and high capacity

\textbf{Keywords:} Charged Qubit, Bloch vectors Entanglement,Capacity .%
\textbf{\ } 

\section{Introduction}

Quantum entanglement is a quantum mechanical phenomenon in which the
quantum states of two or more objects have to be described with
reference to each other, even though the individual objects may be
spatially separated. It is an essential component in many quantum
information processing applications, such as quantum computation
\cite{DiV}, teleportation \cite{Ben} and cryptography \cite{Ekert}
and dense coding \cite{Ben1,San}. Therefore, it is essential to
create and manipulate entangled states for quantum information
tasks. The basic element of the quantum information is the quantum
bit (qubit) which is considered as a two level system. Consequently,
most of the research concentrate to generate entanglement between
two level systems. Among these systems, superconducting charged
qubits i.e. the Cooper pair box \cite{Zhang,Tod}.

In the last years the generated entangled states between Cooper pairs has
been received a lot of attenuation. This is due to its properties as a
two-level quantum system, which makes it a candidate as a qubit in a quantum
computer \cite{Mak, Wend}. Also, recently these types of charged particles
has been used to implement a Shor's factorization \cite{Virt}, Deutsch-Jozsa
algorithm \cite{Faz}, the Grover search algorithm protocol \cite{ Ping}, and
the quantum computing with a single molecular ensemble. On the other hand
theses charged pairs have been used to realize a controlled phase gate \cite%
{Sch}. All these applications have encouraged different authors to study the
entangled properties of these systems. As an example, the entanglement
between two superconducting qubits has been generated by the interaction
with nonclassical radiation \cite{Pater}. Rodrigues et al. have evaluated
the entanglement of superconducting charge qubits by homodyne measurement
\cite{Rod,Rod1}. The entropy squeezing and the emission spectra of a
single-cooper pair have been investigated \cite{Aty}.

On the other hand, it is known that the Bloch vector gives one of the
possible descriptions of N-level quantum states, because it is defined as a
vector whose components are expectation values of some observables. So, we
can use it to describe the density operator of any system \cite{Jak}. These
vectors (called coherent vectors), play a central role in quantifying the
degree of entanglement, where Englert \cite{Englert} has used it to evaluate
the entanglement dyadic which represent a measure of the degree of
entanglement. Addressing this issue with Cooper pair box systems is the main
aim of the present paper.

In this article we propose to study a system of two-Cooper pairs
interacting with a single cavity mode. We investigate the dynamics
of the polarized vector for each individual qubit and the channel
capacity of the entangled Cooper pair for different values of the
system parameters. This paper is organized as follows. In section
$2$, we introduce our model and its solution. We devote section 3,
to investigate the relation between the Bloch vectors and the
quantum entanglement. In section 4, we investigate the effect of
the structure of the initial state and the mean photon number on
the transmission rate of information between two parties via the
channel capacity. Finally, we end up with our conclusion.

\section{Model and its solution}

Our system consists of two superconducting charge qubits each
coupled capacitively to a strapline resonator. Each charged qubit
consists of a small superconducting island with a Cooper pair
charge $\mathcal{Q}$. This island connected by two identical
Josephson junctions, with capacity $C_j$ and Josephson coupling
energy $E_j$, to a superconducting electrode. This system can be
describe as a pair of two-level systems coupled to a simple
harmonic oscillator. The charging energy of the qubits and their
coupling to the resonator can be controlled by the application of
magnetic and electric fields. By using the a rotating wave
approximation, the system can be described by the Hamiltonian
\cite{Rod}.
\begin{equation}
H=\hbar\omega(a^\dagger a+\frac{1}{2})+(E_1\sigma_z+E_2\tau_z)+
\hbar \sum_{i=1}^{2}{\lambda_i(a\sigma^+_i+a^\dagger\sigma^-)},
\end{equation}
where $\sigma_z=\bigl| e \bigr\rangle_1\bigl\langle e \bigr|-\bigl| g %
\bigr\rangle_1\bigl\langle g \bigr|$ for the first qubit,$\tau_z=\bigl| e %
\bigr\rangle_2\bigl\langle e \bigr|-\bigl| g \bigr\rangle_2\bigl\langle g %
\bigr|$ for the second qubit, $a^\dagger$ and $a$ are the creation and
annihilation operators of photons with frequency $\omega$, $E_{1,2}$ are the
charging energies of the qubits, and $\lambda_{1,2}$ are the resonator-qubit
coupling terms.

Our target of this section is to derive the time-dependent density matrix
which enables us to discuss the statistical properties of the present model.
For this reason let us assume that the initial state of the two charged
qubits are prepared in a superposition state which can be written as $\bigl|%
\psi _{c}(0)\bigr\rangle=a\bigl|ee\bigr\rangle+b\bigl|gg\bigr\rangle%
,|a|^{2}|+|b|^{2}=1$, while the cavity field starts from a
coherent state, $\ket{\psi _{f}(0)} =\sum_{n=0}^{\infty
}q_{n}|n\rangle ,~q_{n}=\frac{\alpha ^{n}}{\sqrt{n!}}\exp
(-\frac{1}{2}|\alpha |^{2})$. Now we can write the time evolution
of the density operator in the form
\begin{equation}
\rho (t)=\mathcal{U(}t\mathcal{)}\rho (0)\mathcal{U}^{\dag }\mathcal{(}t%
\mathcal{)},
\end{equation}%
where $\mathcal{U}(t)=\exp \left( -iHt/\hslash \right) $ is the
time-dependent unitary operator and $\rho (0)=|\psi (0)\rangle \langle \psi
(0)|$. Since the invariant sub-space of the global system can be considered
as a set of complete basis of the atom-field. Since, we are interested in
the contribution of the Bloch vector in quantifying the degree of
entanglement, we express the density operator$\rho _{c}(t)$ of the system
charged system by means of the Bloch vector for each qubit and the cross
dyadic. After tracing out the state of the field one gets the state of the
charged qubit as,
\begin{equation}
\rho _{c}(t)=\frac{1}{4}(1+\mathord{\buildrel{\lower3pt\hbox{$%
\scriptscriptstyle\rightarrow$}}\over s}\cdot {\sigma ^{\raisebox{2pt}[%
\height]{$\scriptstyle\downarrow$}}}+\mathord{\buildrel{\lower3pt\hbox{$%
\scriptscriptstyle\rightarrow$}}\over t}\cdot {\tau ^{\raisebox{2pt}[%
\height]{$\scriptstyle\downarrow$}}}+\mathord{\buildrel{\lower3pt\hbox{$%
\scriptscriptstyle\rightarrow$}}\over \sigma}\cdot \mathord{\dyadic@rrow{C}}%
\cdot {\tau ^{\raisebox{2pt}[\height]{$\scriptstyle\downarrow$}}}),
\end{equation}%
where ${\sigma ^{\raisebox{2pt}[\height]{$\scriptstyle\downarrow$}}}=(\sigma
_{x},\sigma _{y},\sigma _{z})$, ${\tau ^{\raisebox{2pt}[\height]{$%
\scriptstyle\downarrow$}}}=(\tau _{x},\tau _{y},\tau _{z})$are the Pauli
matrices for the first and second qubit respectively, while $%
\mathord{\buildrel{\lower3pt\hbox{$\scriptscriptstyle\rightarrow$}}\over s}%
=(s_{x},s_{y},s_{z}),\mathord{\buildrel{\lower3pt\hbox{$\scriptscriptstyle%
\rightarrow$}}\over t}=(t_{x},t_{y},t_{z})$ and are the Bloch vectors for
the first and the second qubits and $\mathord{\dyadic@rrow{C}}$ is the cross
dyadic, are defined as follows:
\begin{eqnarray}
s_{x}(t) &=&A_{n+1}(t)c_{n}^{\ast }(t)+B_{n+1}(t)D_{n}^{\ast
}(t)+C_{n}(t)A_{n+1}^{\ast }(t)+D_{n}(t)B_{n+1}^{\ast }(t),  \nonumber
\label{eqs} \\
s_{y}(t) &=&i\left[ -A_{n+1}(t)C_{n}^{\ast }(t)-B_{n+1}(t)D_{n}^{\ast
}(t)+C_{n}(t)A_{n+1}^{\ast }(t)+D_{n}(t)B_{n+1}^{\ast }(t)\right] ,
\nonumber \\
s_{z}(t) &=&|A_{n}(t)|^{2}+|B_{n}(t)|^{2}-|C_{n}(t)|^{2}-|D_{n}(t)|^{2},
\end{eqnarray}%
while for the second qubit,
\begin{eqnarray}
t_{x}(t) &=&A_{n+1}(t)B_{n}^{\ast }(t)+B_{n}(t)A_{n+1}^{\ast
}(t)+C_{n+1}(t)D_{n}^{\ast }(t)+D_{n}(t)C_{n+1}^{\ast }(t),  \nonumber
\label{eqt} \\
t_{y}(t) &=&i\left[ -A_{n+1}(t)B_{n}^{\ast }(t)+B_{n}(t)A_{n+1}^{\ast
}(t)-C_{n+1}(t)D_{n}^{\ast }(t)+D_{n}(t)C_{n+1}^{\ast }(t)\right] ,
\nonumber \\
t_{z}(t) &=&|A_{n}(t)|^{2}-|B_{n}(t)|^{2}+|C_{n}(t)|^{2}-|D_{n}(t)|^{2},
\end{eqnarray}%
In addition to the elements of the cross dyadic, $\mathord{\dyadic@rrow{C}}$
\begin{eqnarray}
c_{xx} &=&A_{n+2}(t)D_{n}^{\ast }(t)+B_{n}(t)C_{n}^{\ast
}(t)+C_{n}(t)B_{n}^{\ast }(t)+D_{n}(t)A_{n+2}^{\ast }(t),  \nonumber
\label{eq C} \\
c_{xy} &=&i\left[ -A_{n+2}(t)D_{n}^{\ast }(t)+B_{n}(t)C_{n}^{\ast
}(t)-C_{n}(t)B_{n}^{\ast }(t)+D_{n}(t)A_{n+2}^{\ast }(t)\right] ,  \nonumber
\\
c_{xz} &=&A_{n+1}(t)C_{n}^{\ast }(t)-B_{n+1}(t)D_{n}^{\ast
}(t)+C_{n}(t)A_{n+1}^{\ast }(t)-D_{n}(t)B_{n+1}^{\ast }(t)  \nonumber \\
c_{yx} &=&i\left[ -A_{n+2}(t)D_{n}^{\ast }(t)-B_{n}(t)C_{n}^{\ast
}(t)+C_{n}(t)B_{n}^{\ast }(t)+D_{n}(t)A_{n+2}^{\ast }(t)\right] ,  \nonumber
\\
c_{yy} &=&-A_{n+2}(t)D_{n}^{\ast }(t)+B_{n}(t)C_{n}^{\ast
}(t)+C_{n}(t)B_{n}^{\ast }(t)-D_{n}(t)A_{n+2}^{\ast }(t),  \nonumber \\
c_{yz} &=&i\left[ -A_{n+1}(t)C_{n}^{\ast }(t)+B_{n+1}(t)D_{n}^{\ast
}(t)+C_{n}(t)A_{n+1}^{\ast }(t)-D_{n}(t)B_{n+1}^{\ast }(t)\right] ,
\nonumber \\
c_{zx} &=&A_{n+1}(t)B_{n}^{\ast }(t)+B_{n}(t)A_{n+1}^{\ast
}(t)-C_{n+1}(t)D_{n}^{\ast }(t)-D_{n}(t)C_{n+1}^{\ast }(t),  \nonumber \\
c_{zy} &=&i\left[ -A_{n+1}(t)B_{n}^{\ast }(t)+B_{n}(t)A_{n+1}^{\ast
}(t)-C_{n+1}(t)D_{n}^{\ast }(t)+D_{n}(t)C_{n+1}^{\ast }(t)\right]   \nonumber
\\
c_{zz} &=&|A_{n}(t)|^{2}-|B_{n}(t)|^{2}-|C_{n}(t)|^{2}+|D_{n}(t)|^{2},
\end{eqnarray}%
where,
\begin{eqnarray}
A_{n}(t) &=&\sum_{n}^{\infty }\Bigl\{\frac{-a}{\mu _{n}-\nu _{n}}\Bigl(%
(1+R^{2})\beta _{n}^{2}(\cos t\sqrt{\mu _{n}})-\cos (t\sqrt{\nu _{n}}))+\mu
_{n}\cos (t\sqrt{\mu _{n}})+\nu _{n}\cos (t\sqrt{\nu _{n}})\Bigr)  \nonumber
\\
&+&\frac{2b\beta _{n}\gamma _{n}}{\mu _{n}-\nu _{n}}\bigl(\cos (t\sqrt{\mu
_{n}})-\cos (t\sqrt{\nu _{n}})\bigr)\Bigr\},  \nonumber \\
B_{n}(t) &=&\sum_{n}^{\infty }\Bigl\{\frac{-iaR\gamma _{n}}{\mu _{n}-\nu _{n}%
}\Bigl(\frac{(1-R^{2})\beta _{n}^{2}+\mu _{n}}{\sqrt{\mu _{n}}}\sin (t\sqrt{%
\mu _{n}})+\frac{(1-R^{2})\beta _{n}^{2}-\nu _{n}}{\sqrt{\nu _{n}}}\sin (t%
\sqrt{\nu _{n}}~)\Big)  \nonumber \\
&&+\frac{ib\beta _{n}}{\mu _{n}-\nu _{n}}\Bigl(((1-R^{2})\gamma _{n}^{2}-\mu
_{n})\sin (t\sqrt{\mu _{n}})-((1-R^{2})\gamma _{n}^{2}-\nu _{n})\sin (t\sqrt{%
\nu _{n}}~)\Biggr)\Bigl\},  \nonumber \\
C_{n}(t) &=&\sum_{n=0}^{\infty }\Bigl\{\frac{ia\gamma _{n}}{\mu _{n}-\nu _{n}%
}\Bigl(\frac{(1-R^{2})\beta _{n}^{2}-\mu _{n}}{\sqrt{\mu _{n}}}\sin (t\sqrt{%
\mu _{n}})-\frac{(1-R^{2})\beta _{n}^{2}-\nu _{n}}{\sqrt{\nu _{n}}}\sin (t%
\sqrt{\nu _{n}}~)\Biggr)  \nonumber \\
&-&\frac{ibR\beta _{n}}{\mu _{n}-\nu _{n}}\Bigl(\frac{(1-R^{2})\gamma
_{n}^{2}+\mu _{n}}{\sqrt{\mu _{n}}}\sin (t\sqrt{\mu _{n}})+\frac{%
(1-R^{2})\gamma _{n}^{2}-\nu _{n}}{\sqrt{\nu _{n}}}\sin (t\sqrt{\nu _{n}})%
\Biggr)\Bigl\},  \nonumber \\
D_{n}(t) &=&\sum_{n=0}^{\infty }\Bigl\{-\frac{b}{\mu _{n}-\nu _{n}}\Bigl(%
(1+R^{2})\gamma _{n}^{2}(\cos (t\sqrt{\mu _{n}})-\cos (t\sqrt{\nu _{n}}%
))+\mu _{n}\cos (t\sqrt{\mu _{n}})+\nu _{n}\cos (t\sqrt{\nu _{n}})\Bigr)
\nonumber \\
&+&\frac{2a\beta _{n}\gamma _{n}}{\mu _{n}-\nu _{n}}\Bigl(\cos (t\sqrt{\mu
_{n}})-\cos (t\sqrt{\nu _{n}})\Bigr)\Bigl\},
\end{eqnarray}%
and, $\gamma _{n}=\sqrt{1+n},~\quad \beta _{n}=\sqrt{n+2}$ $\mu _{n}=\frac{1%
}{2}(\delta _{n}+\sqrt{\delta _{n}^{2}-4\Delta _{n}^{2}})$, $\nu _{n}=\frac{1%
}{2}(\delta _{n}-\sqrt{\delta _{n}^{2}-4\Delta _{n}^{2}})$ $\delta
_{n}=(1+R^{2})(\gamma _{n}^{2}+\beta _{n}^{2})$ , $\Delta
_{n}=(1-R^{2})^{2}\beta _{n}^{2}\gamma _{n}^{2}$ and $R=\frac{\lambda _{2}}{%
\lambda _{1}}$

\section{Dynamics of Bloch Vectors}

In this section, taking into account the possible extension of the
methods which will be discussed in this work to higher-dimensional
situations and to mixed states, the present strategy will certainly
need to be modified and integrated by developing more sophisticated
tools, including the use of multiple operations to yield properly
defined entanglement and generalizations to mixed-states cases. At
this end, we can say that, the qubit
$\bigl| \psi \bigr\rangle=a\bigl| 0 \bigr\rangle+b\bigl| 1 %
\bigr\rangle$, can be represented as a point $(\theta,\phi)$ on a
unit sphere called the Bloch sphere. Define the angles $\theta $ and
$\phi$ by letting $a=\cos(\frac{\theta}{2})$ and
$b=e^{i\phi}\sin(\frac{\theta}{2})$, then $\bigl| \psi \bigr\rangle$ is represented by the unit vector $%
(\cos\phi\sin\theta,\sin\phi\sin\theta,\cos\theta)$ called Bloch vector.

Here, we investigate the dynamics of the Bloch vectors (Polarized
vector), for different values of the mean photon number $\bar{n}$,
the relative coupling $R$ and different classes of the initial state
of the charged system. In Fig. (1), we assume that the charged
qubits are prepared in the excited state
$\rho_{e}(0)=\bigl|ee\bigr\rangle\bigl\langle ee\bigr|$ and the mean
photon number $\bar{n}=20$, while the relative coupling $R=\lambda
_{2}/\lambda _{1}$ takes different values. In Fig. (1a), we consider
a small value of $R$, which means that the coupling between the
second  qubit and the cavity field is very weak.
\begin{figure}[tbp]
\begin{center}
\includegraphics[width=18pc,height=12pc]{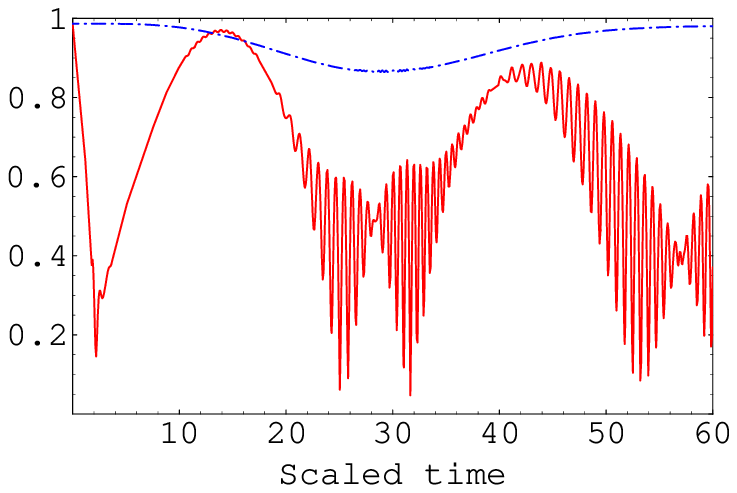}\put(-35,125){(a)} %
\includegraphics[width=18pc,height=12pc]{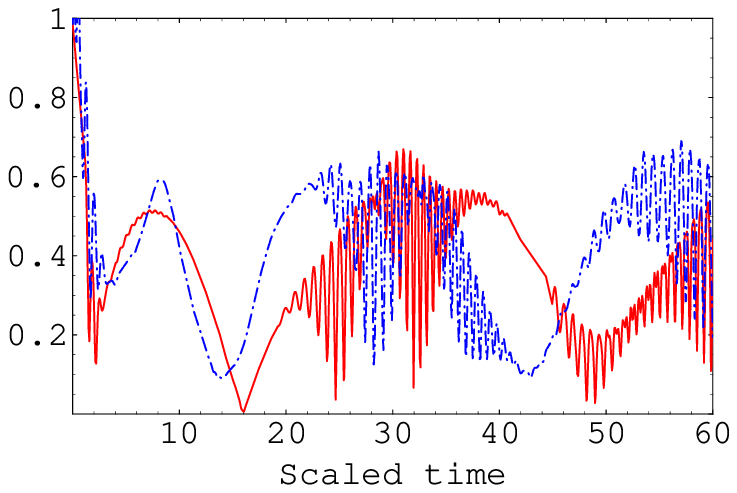}\ \put(-35,125){(b)}\
\end{center}
\caption{ The Bloch vectors as functions of the scaled time. The
two-qubit system is prepared initially in an excited state
$|e_1,e_2\rangle$, and the field starts from a coherent state with
a mean-photon number $\bar{n}=20$. The solid curves for the first
qubit while the dot curves for the second qubit. The relative
coupling is (a) for $R=0.003 $ (b) $R=0.9$.}
\end{figure}
It is clear that, the Bloch vector for the second qubit is bigger
than the Bloch vector for the first qubit. This is due to the weak
coupling between the second qubit and the cavity field. So, the
chance that the second qubit interacts with the field is very small.
On the other hand, as soon as the interaction goes on, we can see
that the Bloch vector for the first qubit decreases rapidly. This is
due to the strong coupling  between the first charged qubit and the
cavity mode. So for this situation, there is an entangled state
generated between the first qubit and the filed, while the second
qubit is almost factorized.
 This phenomenon is not observed in Fig. $(1b)$, where we consider large
values of the coupling $R$. For the numerical values we consider
$R=0.9$, i.e $\lambda_1\simeq\lambda_2$. In this case the two
Bloch  vectors increase and decrease together. This means that the
probability that  both qubits  interact with the field is almost
the  same. Also, we can notice that, this choice  causes a shrink
of the Bloch vectors, where the maximum values almost $0.6$.

It is interesting to mentioned to the fact that, the maximum
entangled state as Bell states has zero Bloch vectors. So, as the
Bloch vectors decrease, the possibility for obtaining an entangled
state with high degree of entanglement increases. So, by using a
suitable values of the coupling constant, one can control the
Bloch vectors and consequently it is possible to generate an
entangled state with high degree of entanglement.
\begin{figure}
\begin{center}
\includegraphics[width=8pc,height=9pc]{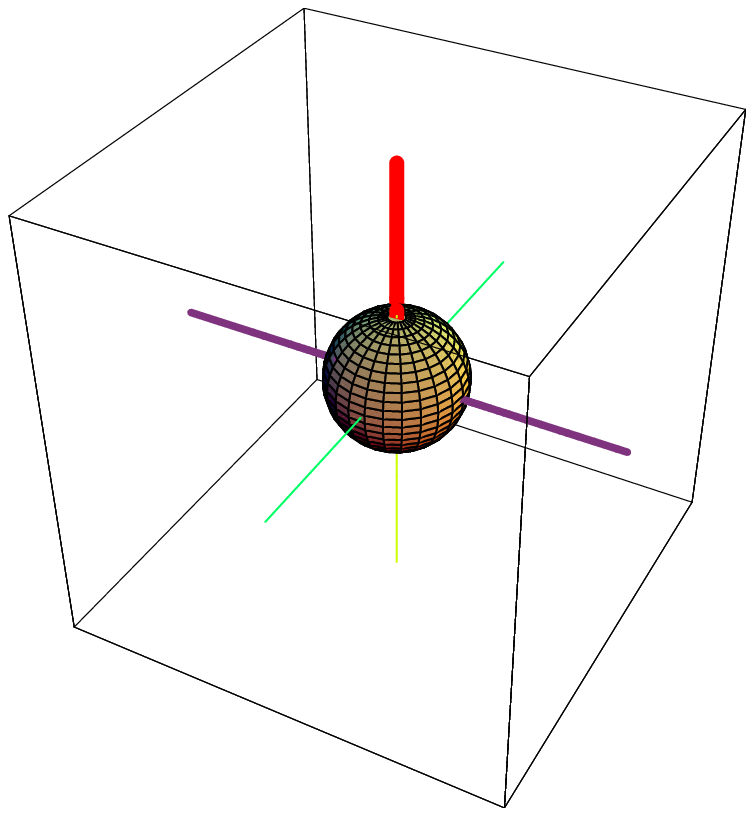}\ \put(-40,110){$%
\lambda t=10.2$}
\includegraphics[width=8pc,height=9pc]{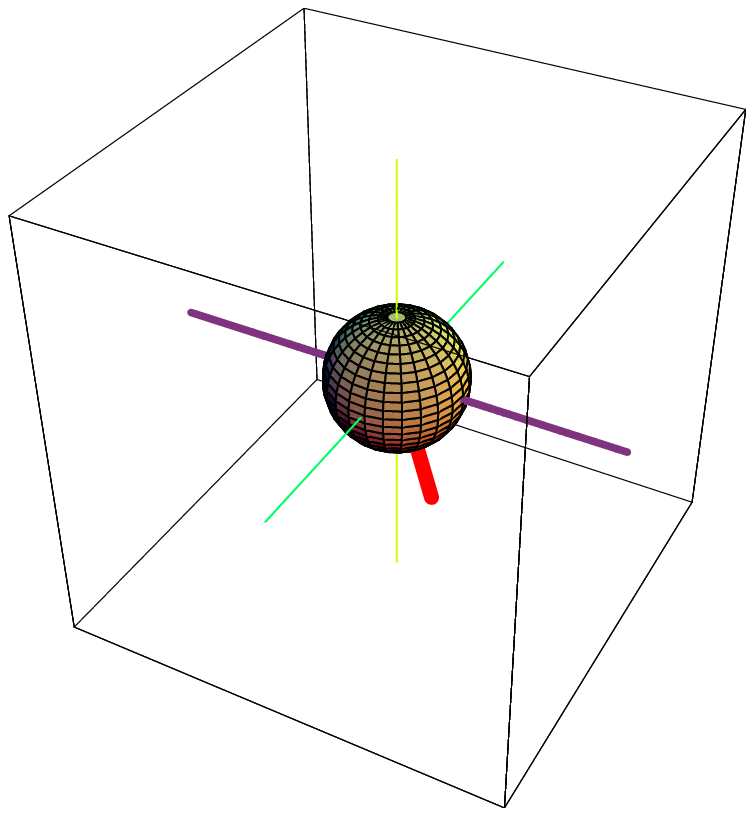}\
\put(-35,110){$\lambda t=10.5$} %
\includegraphics[width=8pc,height=9pc]{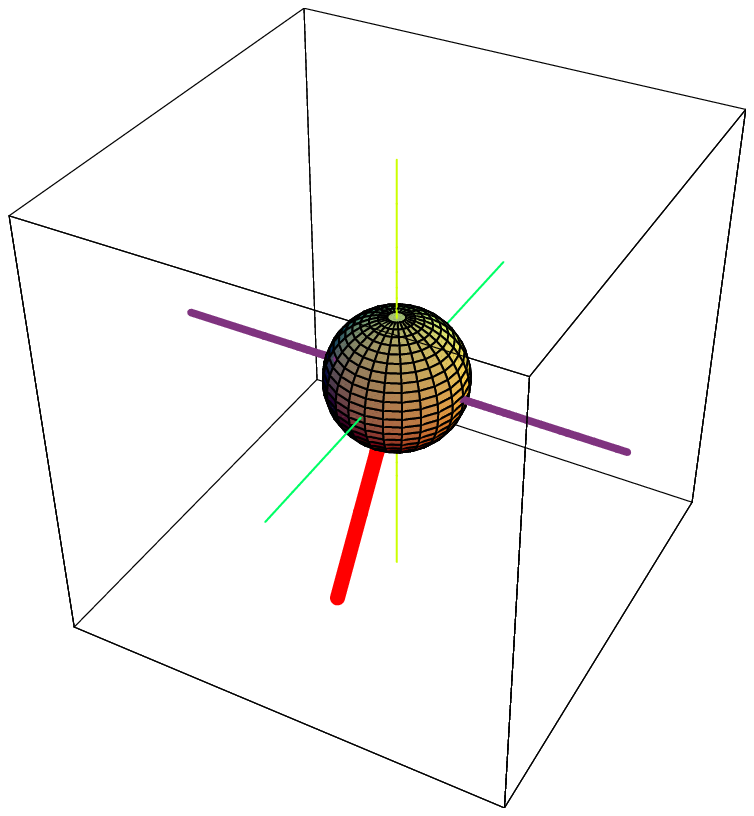}\ \put(-35,110){$%
\lambda t=10.6$}
\includegraphics[width=8pc,height=9pc]{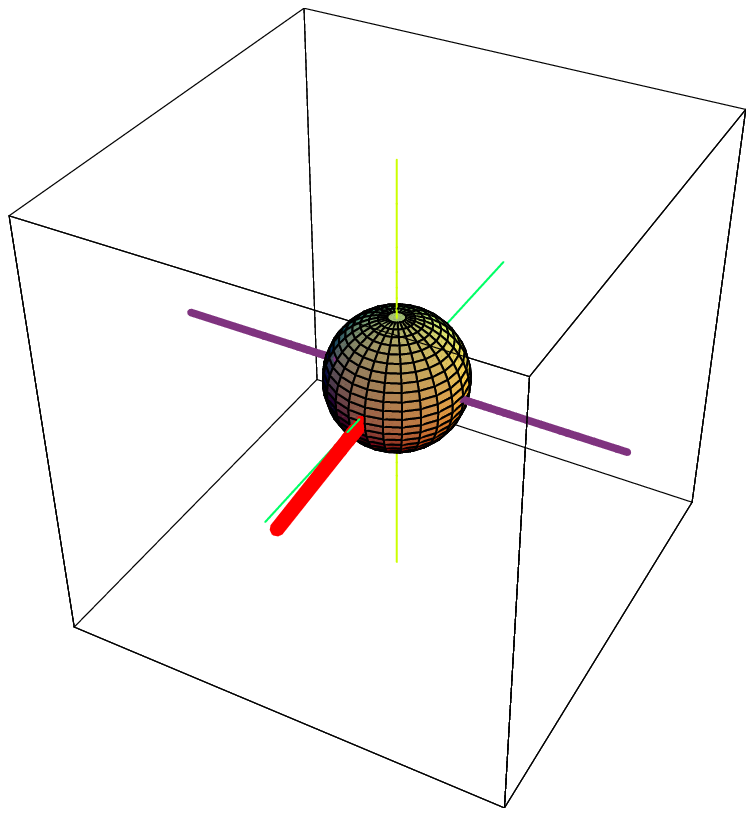}\
\put(-35,110){$\lambda t=10.7$}\ %
\includegraphics[width=8pc,height=9pc]{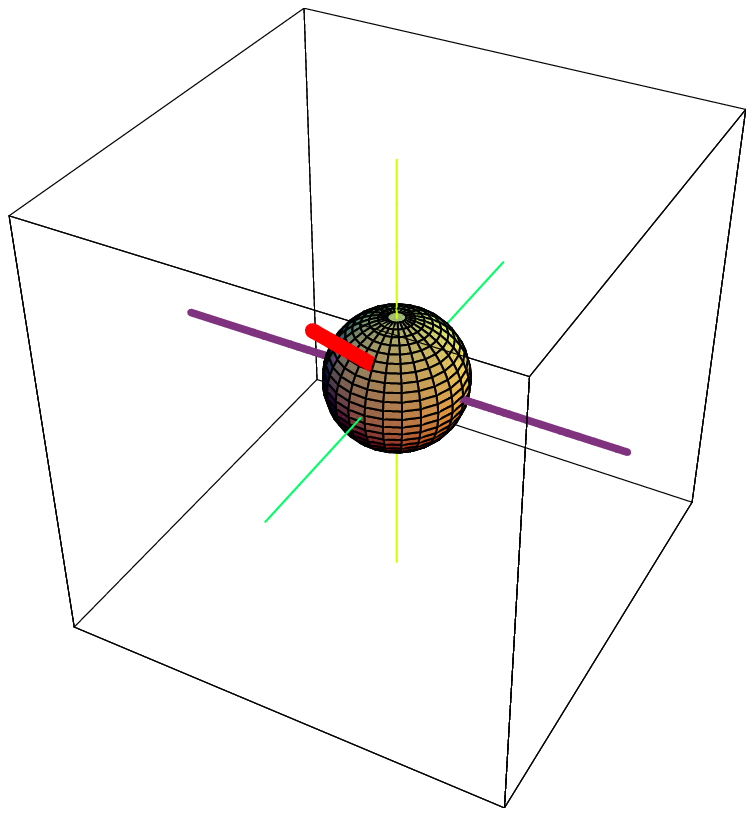}\ \put(-35,108){$%
\lambda t=10.8$}
\includegraphics[width=8pc,height=9pc]{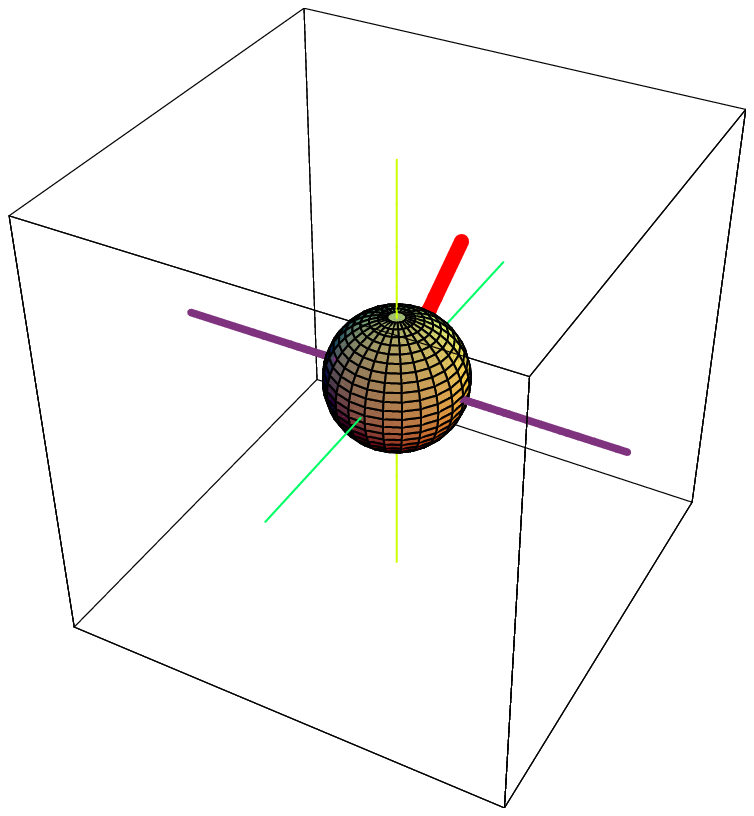}\
\put(-35,108){$\lambda t=11$} %
\includegraphics[width=8pc,height=9pc]{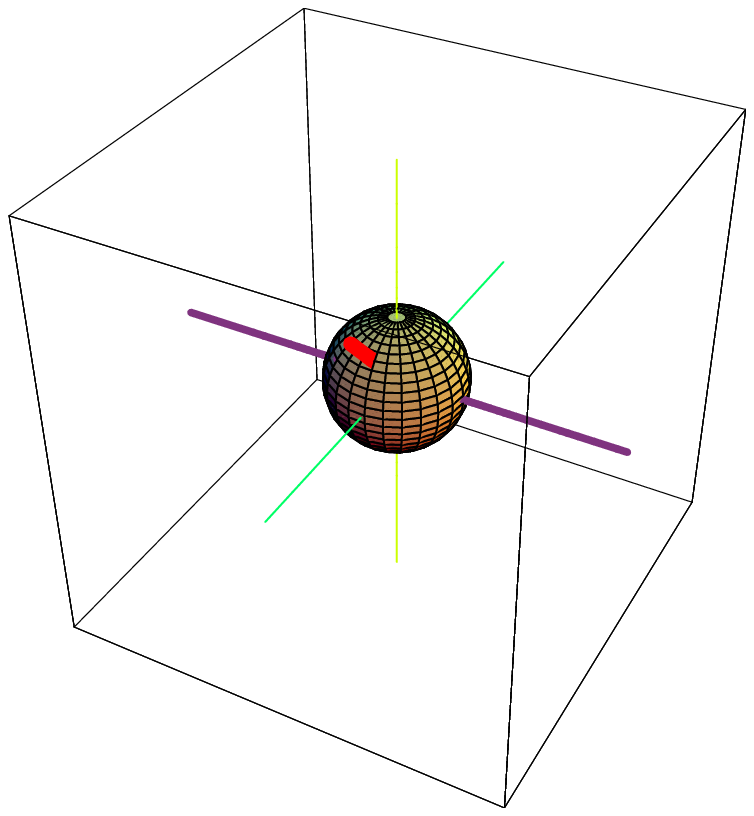}\ \put(-35,108){$%
\lambda t=11.5$}
\includegraphics[width=8pc,height=9pc]{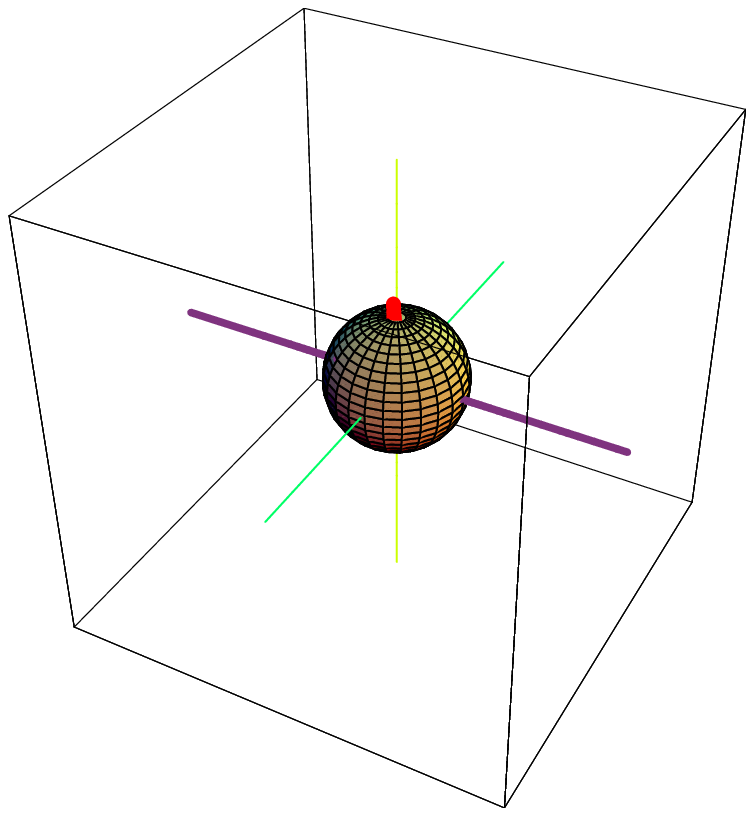}\
\put(-35,108){$\lambda t=11.6$}
\end{center}
\caption{ The Bloch sphere for a charged system prepared initially
in an excited state with $\bar n=20, R=0.9$. The thick line
represents the length of the Bloch vector for the first qubits. }
\end{figure}
In Fig. $(2)$, we plot the Bloch sphere, with radius equal $0.4$
units, at specific time in the interval $[10.2-11.6]$, where we
consider that the charged qubits are prepared in the excited state
and $R=0.9$. The amplitude and the direction of the Bloch vector for
the first qubit are shown. One can see that at $t=10.2$, the Bloch
vector parallel to $y$-axis and it has a large value. As the time
increases (say $t=10.3,10.4$), the Bloch vector inside the sphere
and it could great or equal $0.4$. As the time increases further,
the Bloch vector appears in a different direction and its amplitude
shrinks, this is shown for $t=10.5$. As time increases more the
length of the Bloch vector increases and its direction change.

\begin{figure}[b!]
\begin{center}
\includegraphics[width=17pc,height=12pc]{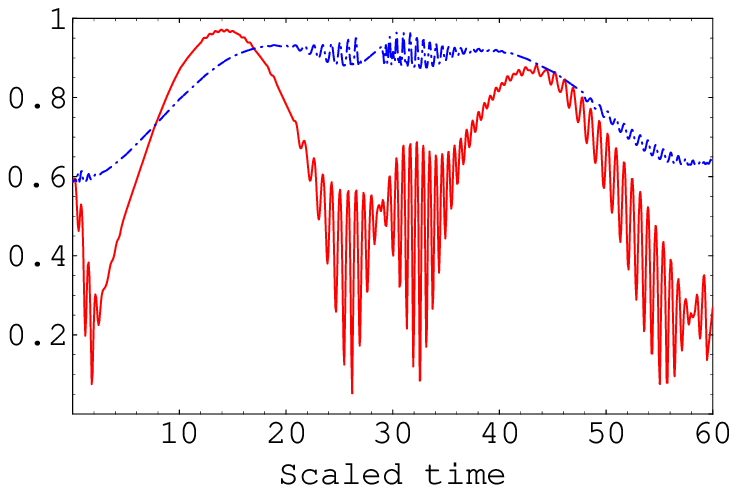}\put(-35,125){(a)} %
\includegraphics[width=17pc,height=12pc]{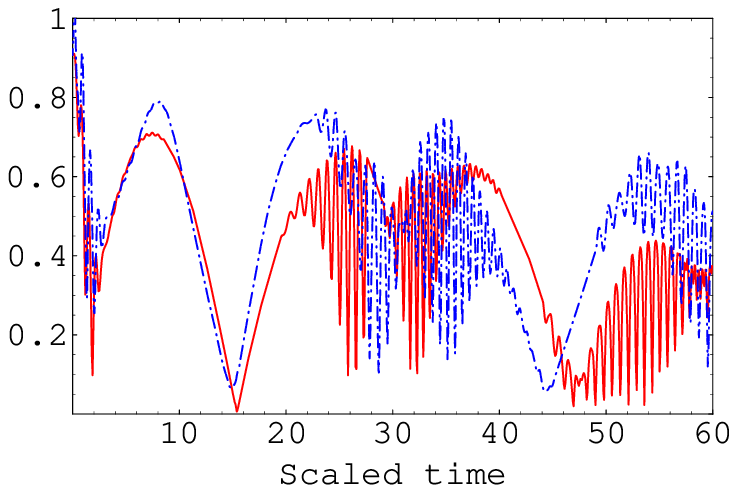}\put(-35,125){(b)}\
\includegraphics[width=17pc,height=12pc]{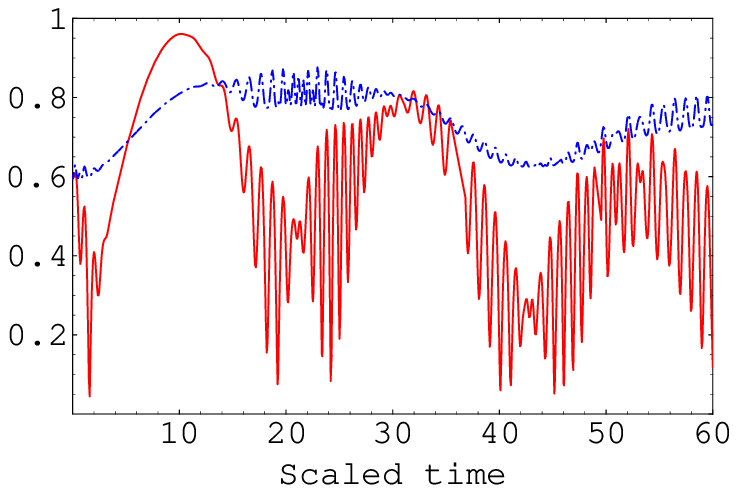}\put(-35,125){(c)}
\includegraphics[width=17pc,height=12pc]{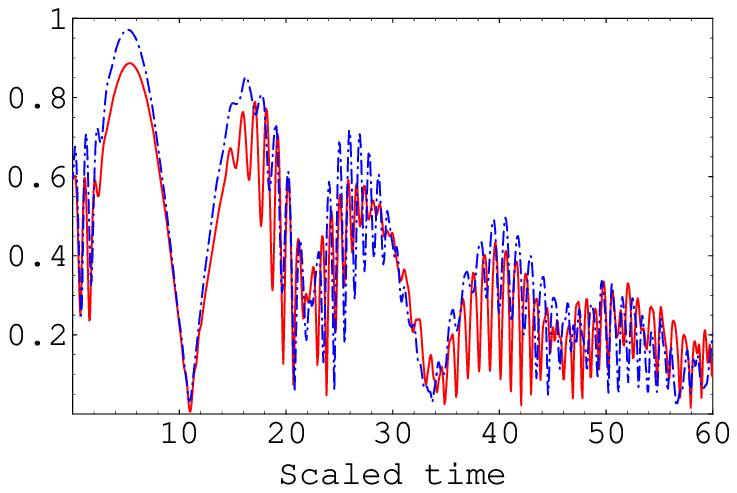}\put(-35,125){(d)}
\end{center}
\caption{The same as Fig.(1) but the charged qubits are prepared initially a
partial entangled state.}
\end{figure}

In Fig. $(3)$, we assume that the charged system is prepared in
the partially entangled state
$a\bigl|ee\bigr\rangle+b\bigl|gg\bigr\rangle$. For small values
for the relative coupling the general behavior of the two vectors
is somehow similar with small differences. Comparing Figs.$(1a)$
and $(2a)$, we see that the minimum point of the Bloch vector is
decreased for the partial entangled state, which means that, using
this class of the initial state, it will be more efficient to
generate entangled state with high degree of entanglement. For
large values of the relative coupling constant, $R=0.9$, the
behavior of the two Bloch vectors are similar. For a specific time
one of the Bloch vectors reaches to zero, in this case there is a
maximum entangled state generated between the two charged qubit,

In Fig. $(3c)$, we consider the mean photon number $\bar n=10$ and
$R$ is small. In this case, we can see that the minimum point of the
Bloch vectors is very small compared with that depicted in Fig.
$(3a)$. Also, the Bloch vector for the second qubit is shrunk more
for small values of the mean photon number. This phenomena is
clearly appeared in Fig. $(3d)$, where we consider  a large value of
the coupling constant.  In additional to the coincidence of the
behavior of the Bloch vector for the two qubits, the two vectors are
shrinking more. So, in this case the possibility of generating
entangled state with high degree of entanglement is increased.

\section{Degree of Entanglement}
To quantify the degree of entanglement between the two charged qubits, we
use a measure which is defined by means of the Bloch vectors and the cross
dyadic. We define the entangled dyadic as
\begin{equation}
\mathord{\dyadic@rrow{E}}=\mathord{\dyadic@rrow{C}}-{s^{\raisebox{2pt}[%
\height]{$\scriptstyle\downarrow$}}}\mathord{\buildrel{\lower3pt\hbox{$%
\scriptscriptstyle\rightarrow$}}\over t}
\end{equation}%
where $\mathord{\dyadic@rrow{c}}$ is a $3\times 3$ matrix which is defined
by(\ref{eq C}) and ${s^{\raisebox{2pt}[\height]{$\scriptstyle\downarrow$}}}%
\mathord{\buildrel{\lower3pt\hbox{$\scriptscriptstyle\rightarrow$}}\over t}$
is also $3\times 3$ matrix whose elements can be obtained from (\ref{eqs})
and (\ref{eqt}). The degree of entanglement is defined by
\begin{equation}\label{DoE}
DoE=tr\{~{\mathord{\dyadic@rrow{E}}}^{\mathsf{T}}\cdot \mathord{%
\dyadic@rrow{E}}\}
\end{equation}%
where ${\mathord{\dyadic@rrow{E}}}^{\mathsf{T}}$ is the transpose matrix of
the dyadic $\mathord{\dyadic@rrow{E}}$
\begin{figure}[tph]
\begin{center}
\includegraphics[width=18pc,height=12pc]{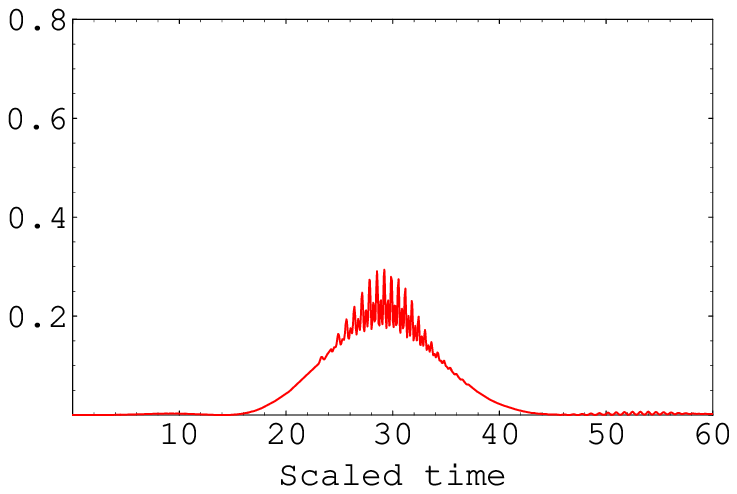}\put(-35,125){(a)}
\includegraphics[width=18pc,height=12pc]{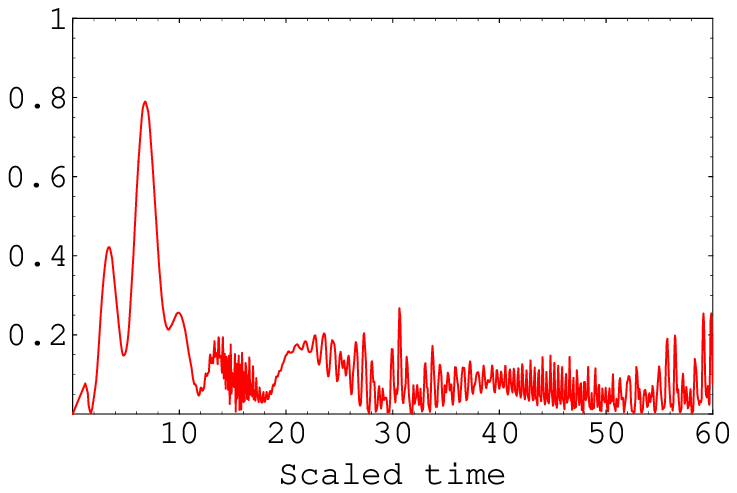}\put(-35,125){(b)}
\end{center}
\caption{ The degree of entanglement for a charged system prepared
initially in an excited state with $\bar{n}=20$. (a) For $R=0.003$
(a) excited (b) $R=0.9$.}
\end{figure}

The amount of entanglement between the entangled charged qubits is
shown in Fig. $4$, in which we consider different regimes. In the
first regime, we assume that the two qubits are prepared initially
in the following excited state $\bigl|\psi
(0)\bigr\rangle_{c}=\bigl|ee\bigr\rangle$ while
the field starts from a coherent state with a mean photon number $\bar{n}=20$%
. In Fig. $(4a)$, we see that for a small value of the coupling
constant (say $R=0.003)$, there is no any quantum correlation
between the two charged qubit expect on the interval [19.2-40.5],
where in this interval the Bloch of the second qubit decreases. This
means that in this intervals, the three subsystems (the two charged
qubits and the field) interact with each others. Also, from this
figure, we can see that the maximum amount of entanglement is
obtained at the minimum point of the Bloch vectors of both qubits.
This phenomena, also, is shown in Fig. $(4b)$, where we consider
$R=0.9$. As soon as the interaction starts, (scaled time is greater
than zero), the entangled state starts to be generated. It is
obvious to realize that the development of the entanglement depends
on the dynamics of the Bloch vectors.
\begin{figure}[b!]
\begin{center}
\includegraphics[width=18pc,height=12pc]{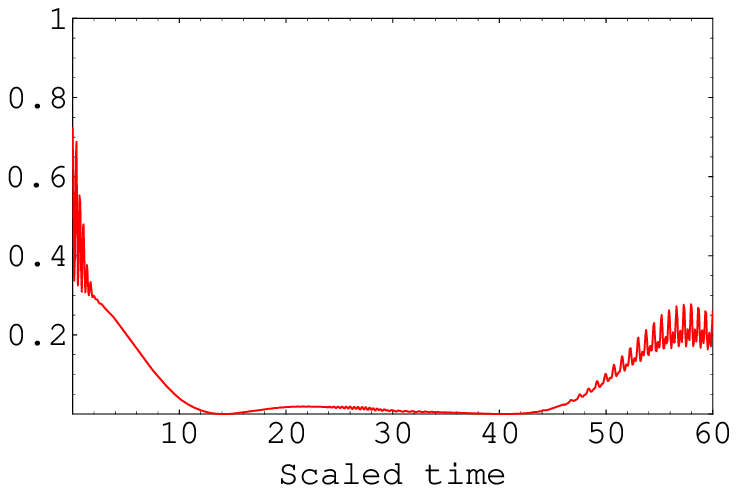}\put(-35,125){(a)}
\includegraphics[width=18pc,height=12pc]{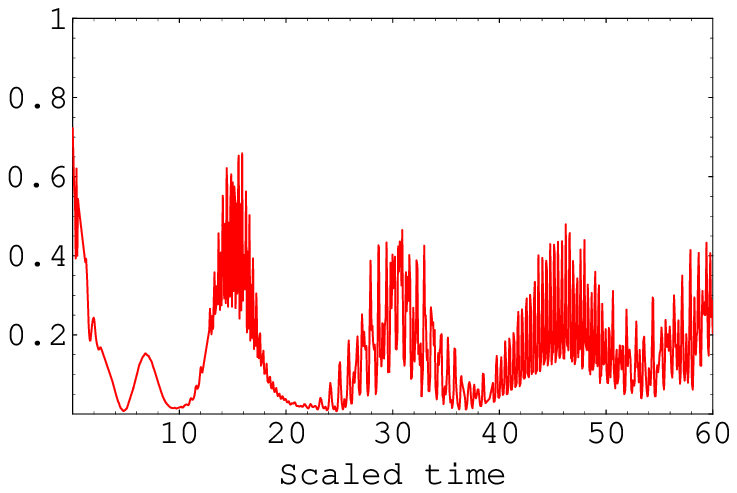}\put(-35,125){(b)}
\end{center}
\caption{ The degree of entanglement for a charged system prepared
initially in partially entangled states with $\bar{n}=20$. (a) For
$R=0.003$ (a) excited (b) $R=0.9$.}
\end{figure}

The effect of different setting of the initial state of the charged
qubit is seen in Fig.(5), where we assume that the system is
prepared initially in a partially entangled state. Fig. $(5a)$, is
devoted to consider a weak  correlation case (say $R=0.003$). From
this figure it is clear that the degree of entanglement decreases
with time until it reaches a minimum point at $\lambda_1 t\simeq
15$. From Fig. $(3a)$, we see that the Bloch vector for both qubit
are maximum and consequently the degree of entanglement will be
minimum according to the definition (\ref{DoE}). In the interval
$[40-60]$, the degree of entanglement starts to increase another
time, where in this interval of time the Bloch vector decrease (see
Fig. $(3a)$). For a strong correlation the dynamics is seen in Fig.
$(5b)$. It is clear that the degree of entanglement is much
stronger, where the efficiency of generating entangled state in this
case is much higher. Also, the evolvement of the degree of
entanglement depends on the evolvement of the Bloch vector for each
qubit.

In this context it is very important to investigate the evolvement
of the degree of entanglement for different values of the mean
photon number $\bar n$. We choose the case for the weak coupling,
where $R=0.003$. In Fig.$(6a)$, we assume that the system  has been
processed in advance in an excites state. The usual effect of the
mean photon number is seen where the Rabi oscillation is shifted to
the left.  In this case the maximum amount of entanglement is
smaller than that depicted for $\bar n=20$~(Fig. $(4a))$. But on the
other hand, there is an entangled state is generated in the interval
$[50-60]$, while on the corresponding interval the charged  system
behaves as a product state. In Fig. $(6b)$, we plot the degree of
entanglement for a charged qubit is prepared in partially entangled
state. In this case the degree of entanglement is much better than
that shown in Fig. $(5a)$, where  $\bar n=20$. So, For the weak
interaction one can generate an entangled state between the two
charged qubits by reducing the number of photons in the cavity mode.
\begin{figure}[t]
\begin{center}
\includegraphics[width=18pc,height=12pc]{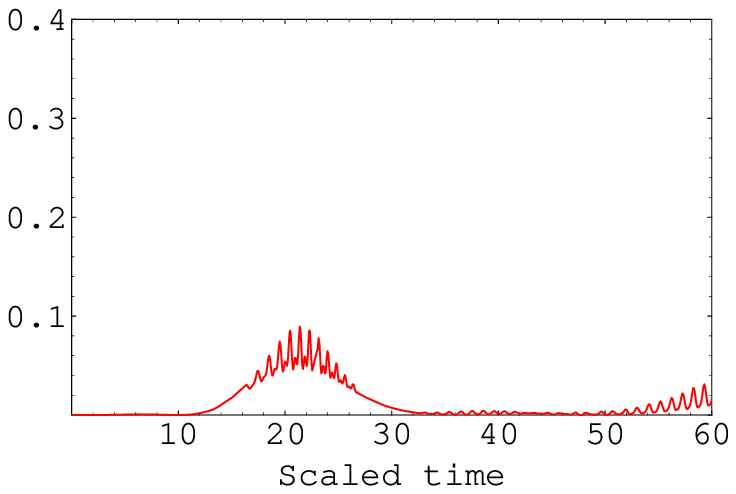}\put(-35,125){(a)}
\includegraphics[width=18pc,height=12pc]{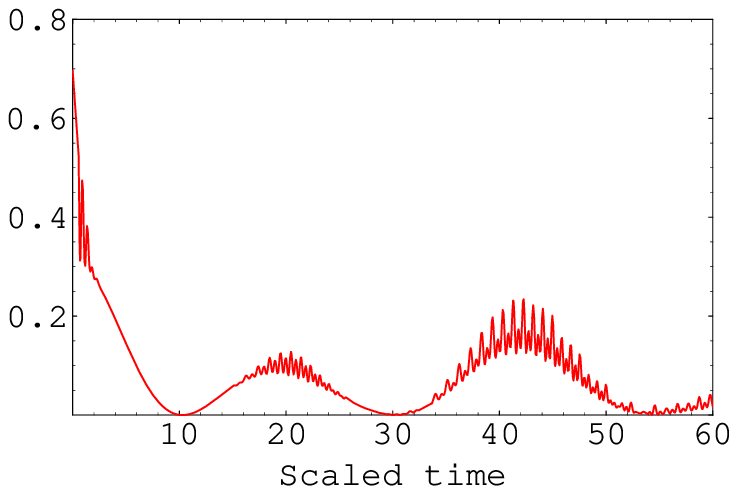}\put(-35,125){(b)}
\end{center}
\caption{ The degree of entanglement for a charged system prepared
initially in partially entangled states with $\bar{n}=10$,
$R=0.003$. For  (a) excited (b) partially entangled state.}
\end{figure}

\section{The channel Capacity}
In this section we investigate the effect of the structure of the
initial state and the mean photon number on the transmission rate
of information from a sender (Alice), to a receiver (Bob). This
task can be performed by employing the dense coding protocol
\cite{Ben}, \cite{bose,Qiu}. The main idea of this protocol is
that, Alice and Bob share an entangled qubit pair. They used it as
a channel, where Alice can encode two classical bits in her qubit
by using local operation. Alice sends her qubit to Bob, who will
try to decode the information. The amount of information gained by
Bob depends on the capacity of the channel. In this context, we
try to show how the capacity of the channel and hence the rate of
transmit information depend on the structure of the initial
states, the relative coupling and the mean photon number. For
$N\times M$ state system, the dense coding capacity is given by
\begin{equation}\label{cap}
\mathcal{C}=log{D_{A}}+\mathcal{S}(\rho _{B})-\mathcal{S}(\rho _{AB}),
\end{equation}%
where $D_{A}=N$, $\rho _{B}=tr_{A}\{{\rho _{AB}}\}$ and
$\mathcal{S}(.)$ is the Von Numann entropy. Since we consider
entangled two qubits, then our system is in $2\times 2$ dimension.
Fig. $5$, shows the behavior of the capacity of the entangled
quantum channel, where we consider the two charged qubits are in
excited state. We investigate the effect of the coupling
constant,the mean photon number and the setting of the initial
state of the charged system.
\begin{figure}[b!]
\begin{center}
\includegraphics[width=17pc,height=12pc]{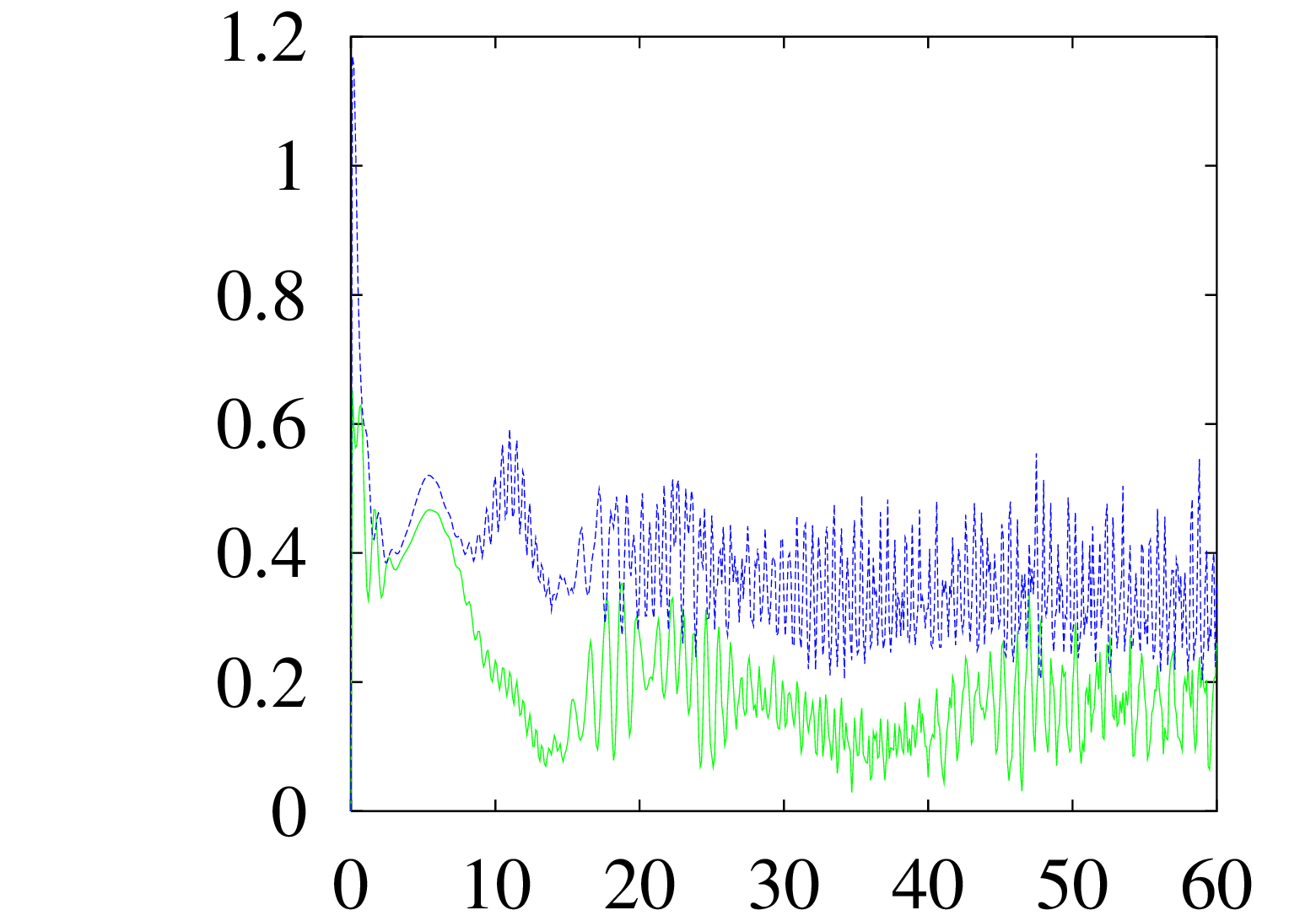}\put(-35,125){(a)}\
\put(-100,-10){Scaled Time}
\includegraphics[width=17pc,height=12pc]{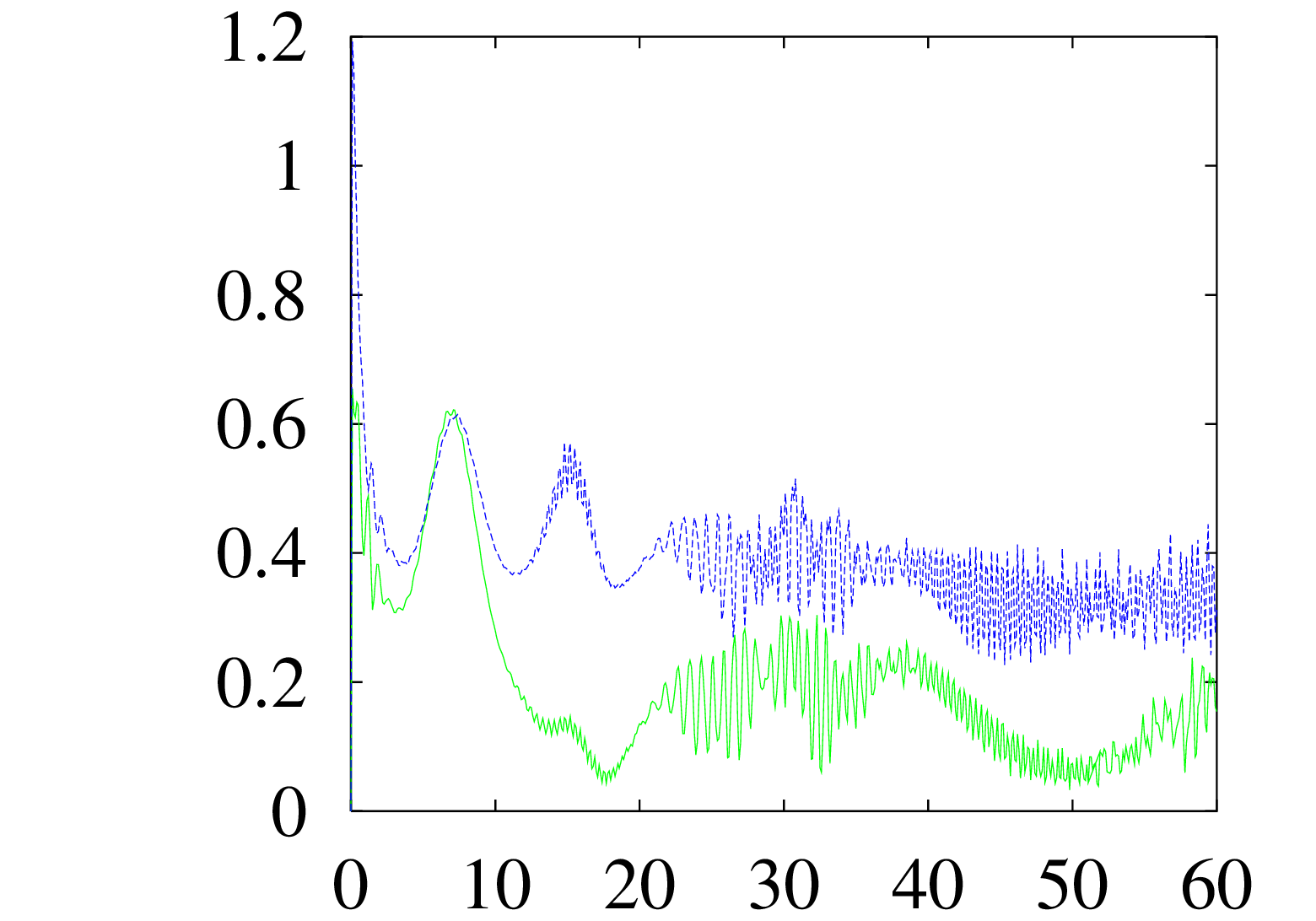}\put(-35,125){(b)}
\put(-100,-10){Scaled Time}
\end{center}
\caption{ The capacity of the channel where $R=0.9$
 The solid  and dot curves for a charged system is prepared
 initially in  a partially entangled and excited states
 respectively. (a) for $\bar n=10$, and (b) For $\bar n=20$. }
\end{figure}
 Fig. $(7a)$, shows the behavior of the channel capacity for
 a charged system with a strong coupling  with the field, where we
 assume that $R=0.9$ and the mean photon number $\bar n=10$. From
 this figure it is clear that the channel capacity for a charged  system is
 prepared initially in  a partially entangled  state is much
 better than that if we start with a charged qubit in an excited
 state (product state). The effect of the mean photon number is
 seen if we take a look at Fig.(7b), where we consider $\bar
 n=10$.  From this figure, we can see that the Rabi oscillations
 shrunk and shifted to the right as time increases. Due to the
 shrunk of the oscillations, we can see that the channel capacity
 increases  a little bit  as $\bar n$ increases.  From Fig.$(4b)$
 and Fig.($7b)$, we can see that there is a strong relation between
 the degree of entanglement and the channel capacity and hence the
 rate of transform data. This maximum value  of the channel
 capacity is obtained at a large degree of entanglement, as an
 example in the interval [0.2,0.8]. Also, from equation
 (\ref{cap}), the capacity of the channel does not only depend
 on the channel $\rho_{ab}$, but also on the individual state for
 the single qubit $\rho_b$. It is clear on the interval [10-20] as an example,
 the degree of entanglement is almost zero, but the capacity  does not vanish.

\section{Conclusion}

In this paper we have studied in a non-standard way the dynamics
of the Bloch vectors for two charged qubits. We investigate the
dynamical behavior of the Bloch vector for each qubit. The shrunk,
extension  and the direction of these vectors are examined for
different parameters of the charged system and the cavity field.
We show for strong coupling between the filed and the charged
system, the Bloch vector for the two qubits have the same
behavior. But if one qubit has a weak coupling with the field (as
the second qubit in our case), the behavior of the two Bloch
vectors is different.

Using the density matrix technique, we predict the existence of
entangled states, where we consider an entanglement measure which
depends on the two Bloch vectors and the cross dyadic and is called
entangled dyadic. The relation between the evolvement of the Bloch
vector and the degree of entanglement is shown, where for large
values of the Bloch vectors, the degree of entanglement is minimum.
It is shown that, for a charged qubit prepared initially in an
entangled state, the amount of entanglement is much larger than that
for any other choices. The role played by the coupling constant and
the mean photon number is cleared for generating entangled states
and improving its degree of entanglement. Decreasing the mean photon
number of the cavity mode is important to generate entangled state
even the coupling between the qubits and field is weak.

The behavior of the channel capacity between the two charged qubits
is examined, using different regimes of preparing the initial state
of the system. It is shown that the possibility of generating
entangled state with high capacity and consequently high rate of
transmission of information is much better if we start with a
partially entangled state for the system. Also, the mean photon
number plays a central role in the efficient of the channel
capacity, where for small values of the mean photon numbers, the
channel capacity is increased and consequently the transmission rate
of information.

\end{document}